# NaInX$_2$ (X = S, Se) layered materials for energy harvesting applications: First-principles insights into optoelectronic and thermoelectric properties

M. M. Hossain[1,*], M. A. Hossain[2,**], S. A. Moon[2], M. A. Ali[1], M. M. Uddin[1], S. H. Naqib[3,***], A. K. M. A. Islam[3,4], M. Nagao[5], S. Watauchi[5] and I. Tanaka[5]

[1]Department of Physics, Chittagong University of Engineering and Technology, Chittagong-4349, Bangladesh
[2]Department of Physics, Mawlana Bhashani Science and Technology University, Santosh, Tangail-1902, Bangladesh
[3]Department of Physics, University of Rajshahi, Rajshahi 6205, Bangladesh
[4]Department of Electrical and Electronic Engineering, International Islamic University Chittagong, Kumira, Chittagong, 4318, Bangladesh
[5]Center for Crystal Science and Technology, University of Yamanashi, 7-32 Miyamae, Kofu, Yamanashi 400-8511, Japan

Corresponding author: *email: mukter_phy @cuet.ac.bd; mukter.phy@gmail.com
**email: anwar647@mbstu.ac.bd
***email: salehnaqib@yahoo.com

## Abstract

In recent times, layered chalcogenide semiconductors have attracted great interest in energy harvesting device applications. In the present study, the structural, electronic, optical and thermoelectric properties of two isostructural chalcogenide materials, NaInS$_2$ and NaInSe$_2$ with hexagonal symmetry (*R*-3*m*) have been studied using the first principles method. A very good agreement has been found between our results with the available experimental and theoretical ones. The studied materials are semiconducting in nature as confirmed from the electronic band structure and optical properties. The strong hybridizations among *s* orbitals of Na, In and Se atoms push the bottom of the conduction band downward resulting in a narrower band gap of NaInSe$_2$ compared to that of NaInS$_2$ compound. Different optical (dielectric function, photoconductivity, absorption coefficient, reflectivity, refractive index and loss function) and thermoelectric (Seebeck coefficient, electrical conductivity, power factor and thermal conductivity) properties of NaInX$_2$ (X = S, Se) have been studied in detail for the first time. It is found that all these properties are significantly anisotropic due to the strongly layered structure of NaInX$_2$ (X = S, Se). Strong optical absorption with sharp peaks is found in the far visible to mid ultraviolet (UV) regions while the reflectivity is low in the UV region for both the compounds. Such features indicate feasibility of applications in optoelectronic sector. The calculated thermoelectric power factors at 1000 K for NaInS$_2$ and NaInSe$_2$ along *a*-axis are found to be 151.5 μW/cmK$^2$ and 154 μW/cmK$^2$, respectively and the corresponding *ZT* values are ~0.70. The obtained thermal conductivity along *a*-axis for both compounds is high (~22 W/mK). This suggests that the reduction of such high thermal conductivity is important to achieve higher *ZT* values of the NaInX$_2$(X = S, Se) compounds.

Keywords: Chalcogenides; electronic properties; optical properties; thermoelectric properties

1. **Introduction**

Ternary chalcogenide semiconductors are considered to be promising materials for various applications such as in photovoltaics, photocatalysts, nonlinear optics, photo-response, and topological insulator applications [1–8]. Among the chalcogenide materials, researchers have paid particular attention to ternary indium compounds like $NaInX_2$ (X = S and Se) chalcogenides due to their fascinating structural features, tunable bandgap and unique stoichiometry-controlled electro-optical properties [3]. Very recently, layered chalcogenides, $NaInS_{2-x}Se_x$ ($x$ = 0, 0.5, 1.0, 1.5 and 2.0) solid solution have been synthesized by Takahashi *et al.*[3] using ball milling process. The study [3] reported that the bandgap of the $NaInS_2$ compound can be tuned by varying Se content.The experimentally estimated band gap of $NaInS_{2-x}Se_x$ ($x$ = 0) is seen to decrease to 2.26 eV when S is replaced completely ($x$ = 1) with Se content. These tunable band gap indicated the potential use of $NaInX_2$ (X = S and Se) as optoelectronic materials in the visible and evennear ultraviolet spectral regions. In addition to this study, $NaInS_2$ compound and its counterparts have also been studied experimentally and/or theoretically by other researchers. Fukuzaki *et al.*[1] have synthesized $NaInX_2$(X = S, O) materials and confirmed the role of anions on the electronic structure using X-ray photoelectron spectroscopy along with first-principles molecular orbital (MO) cluster calculations. The $NaInS_2$, a heterogeneous nanosheet, was synthesized using a partial cation exchange reaction process by P. Hu *et al.*[6]. They reported that the $NaInS_2$ heterogeneous nanosheet shows highly improved photocatalytic behavior.

Recently we have already studied the structural, mechanical and thermodynamic properties in detail of ternary layered chalcogenides, $NaInS_{2-x}Se_x$ (x = 0, 0.5, 1.0, 1.5 and 2.0) solid solutions[9]. In 2020, Yaseen *et al.*[10] have reported interesting results on electronic, optical and thermoelectric properties of body centered tetragonal (BCT) chalcopyrite of $NaInY_2$ (Y = S, Se,Te) with space group *I-42d*. This report shows that the $NaInY_2$ (Y = S, Se, Te) compounds are promising materials for photovoltaic (infrared and visible region) and thermoelectric device applications.This indicates that such study of hexagonal chalcogenide semiconductor $NaInX_2$ (X = S, Se) with space group *R-3m* should be of significant scientific interest. Up to now, most of the experimental and theoretical studies on $NaInX_2$ (X = S, Se) are concerned mainly with the structural and electronic (band structure) properties of the materials. But studies on the total density of states (TDOS) and partial density of states (PDOS), charge density distribution and Mulliken charge population analysis are still unexplored. Optical parameters also remained largely unexplored.The electronic structure of solids can further be understood using the knowledge of energy dependent optical properties [11]. The knowledge of absorption coefficient, refractive indices and dielectric functions are essential to design optoelectronic devices.In addition to possiblephotovoltaic/optoelectronic applications of these compounds, the tunable bandgap and layered crystal structure further indicate the possible use as potential thermoelectric materials.

The coefficient of thermoelectric performance is conventionally measured by the dimensionless figure of merit defined as $ZT = \frac{S^2 \sigma T}{\kappa}$, where $S$, $\sigma$, $\kappa$ (= $\kappa_e + \kappa_l$) and T stand for Seebeck coefficient, electrical conductivity, thermal conductivity and absolute temperature, respectively[12–14]. In general, layered materials exhibit comparatively low thermal conductivity which is essential for achieving high $ZT$ value[14]. To the best of our knowledge, no experimental or theoretical data on optical (except absorption spectra [3]) and thermoelectric properties are available in literature. Therefore, a first-principles study will be performed to investigate electronic, optical and thermoelectric transport properties of hexagonal NaInX$_2$ (X = S, Se) in order to fill this significant research gap and to check the feasibility of theirapplications in optoelectronic and thermoelectric device sectors.

2. **Computational Methodology**

The crystal structures of layered NaInX$_2$ (X=S, Se) compounds have been optimized using CASTEP code [15,16] which is based on the density functional theory (DFT) [17,18].The plane wave basis set cut-off is set to 500 eV, and for the sampling of the Brillouin zone, a 15×15×23 Monkhorst–Pack mesh is employed [19]. Thegeometry optimization is achieved using convergence thresholds of 5×10$^{-6}$ eV/atom for the total energy, 0.01 eV/Å for the maximum force, 0.02 GPa for the maximum stress and 5×10$^{-4}$ Å for maximum atomic displacement. The optimized crystal structures have been used for all calculations in the present study. The electronic band structures were calculated in WIEN2k [20] using the generalized gradient approximation (GGA) within the Perdew Burke-Ernzerhof (PBE)[18,21] scheme and the Tran-Blaha modified Becke-Johnson potential (TB-mBJ) [22]. To obtain a good convergence of the basis set, a plane wave cut off of kinetic energy RK$_{max}$ = 7.0 was selected. A dense mesh of (31×31×5) k-points was used in the electronic, optical and transport properties calculations. The TB-mBJ potential was used for optical and thermoelectric transport properties calculations. The transport properties were calculated by solving semi-classical Boltzmann transport equations as implemented in the widely used

BoltzTraP code [23]. The equation for the transport coefficients are defined in the Boltzmann transport theory [24,25] based on the relaxation time approach.

### 3. Results and discussion

3.1 Structural properties

The chalcogenide compounds, NaInS$_2$ and its counterpart NaInSe$_2$ belong to the hexagonal crystal structure with space group *R-3m* (# No. 166) [1,2]. Fig. 1 shows the two- and three-dimensional crystal structure of the NaInS$_2$ compound. The constructed unit cell has three formula units that contain a total of twelve atoms. The crystallographic lattice parameters and atomic positions as well as the percentage of deviation from earlier reports are tabulated in Table-1 along with available experimental and theoretical results. A maximum discrepancy of the present calculation is found to be 1.70 % for the lattice constant *a*. This deviation of lattice constant from experimental value seems to be fair as the accuracy in theoretical calculations usually depends on the choice of the pseudopotential.

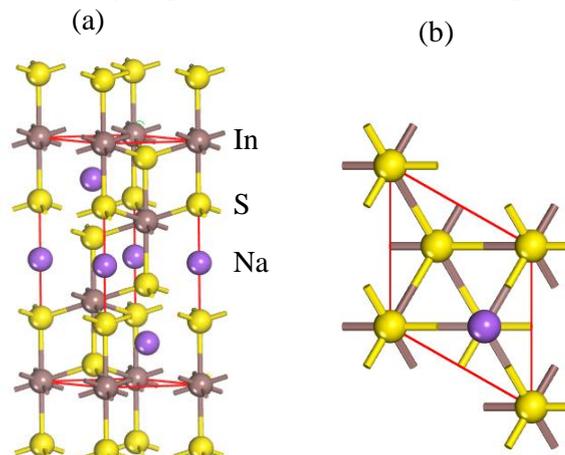

**Fig. 1.** (a) Three dimensional and (b) two dimensional hexagonal layered crystal structure of NaInS$_2$ with space group *R-3m* (# No. 166).

**Table 1-** The crystallographic lattice constants and Wyckoff atomic positions of NaInS$_2$ and NaInSe$_2$ compounds.

| Phases | *a* (Å) | % of deviation | *c* (Å) | % of deviation | Remarks | | Atomic positions | | |
|---|---|---|---|---|---|---|---|---|---|
| | | | | | | | *x* | *y* | *z* |
| NaInS$_2$ | 3.868 | 1.70 | 20.134 | 1.22 | This | Na | 0.0 | 0.0 | 0.5 |
| | 3.803 | | 19.89 | | Expt.[3] | In | 0.0 | 0.0 | 0.0 |
| | | | | | | S/Se | 0.0 | 0.0 | 0.26 |
| NaInSe$_2$ | 4.022 | 1.25 | 21.005 | 0.55 | This | | | | |
| | 3.972 | | 20.890 | | Expt.[3] | | | | |

## 3.2 Electronic properties, Mulliken atomic population and charge density

The previous theoretical study on electronic properties was not adequate in view of the available experimental results [3].Owing to increased interest on prospective materials it is thus necessary to add more information and analyze on electronic properties which may help to design optoelectronic and thermoelectric device applications as well. Since the optical and transport properties strongly depend on electronic band structure of materials, the underestimated band gaps are not suitable to predict optical and transport parameters. A lot of prior studies have reported that GGA-PBE approach underestimate the band gap of semiconducting material while TB-mBJ potential produces band gaps in good agreement with the experimental results [12,26–29]. Figs. 2 (a-d) show the calculated electronic band structures and density of states of $NaInX_2$ (X= S, Se) layered crystals along different symmetry directions in the momentum space using the TB-mBJ potential. It should be mentioned that the estimated band gaps using GGA-PBE potential for the $NaInS_2$ and $NaInSe_2$ are found to be 1.88 and 1.10 eV, respectively, which are in good accord with previous theoretical study for band gap estimation but are very low compared tothe reported experimental values [1,3,5].

In the electronic band structure, the Fermi level ($E_F$) is placed at 0 eV, and a distinct band gap between conduction and valence bands was found. The $E_F$ is located very close to the peak in the valence band. The valence band maximum (VBM) is found at the Γ point for both $NaInS_2$ and $NaInSe_2$, while the conduction band minimum (CBM) is present at the L point and Γ point for $NaInS_2$ and $NaInSe_2$ compounds, respectively.

This implies that the $NaInS_2$ compound is an indirect band gap semiconductor and $NaInSe_2$ possesses a direct bandgap. The TB-mBJ estimated bandgaps are ~ 3.3 and ~2.3 eV for the $NaInS_2$ and $NaInSe_2$ compounds, respectively. These results are very much consistent with the experimental results [1,3]. The lowest part of the conduction band is shifted significantly to lower energy at Γ point while the upper part of the valence band was not noticeably shifted when the S is replaced by Se. This should be the reason for the reduced band gap for the $NaInSe_2$ compound. The total and partial DOS are studied here for the purpose of explaining the contribution from different atomic orbitals to the atomic bonding; charge transport and optical transitions as shown in Figs. 2 (c) and (d). The top of the valence band for both the compounds is mainly comprised of Na-2$p$, S-3$s$, S/Se-$p$ (S for $NaInS_2$ and Se for $NaInSe_2$ compounds), and In-5$p$ and In-4$d$ electronic states. The lower part of valence bands around -4 eV originates from $s$ orbital of Na, In, and S/Se elements.

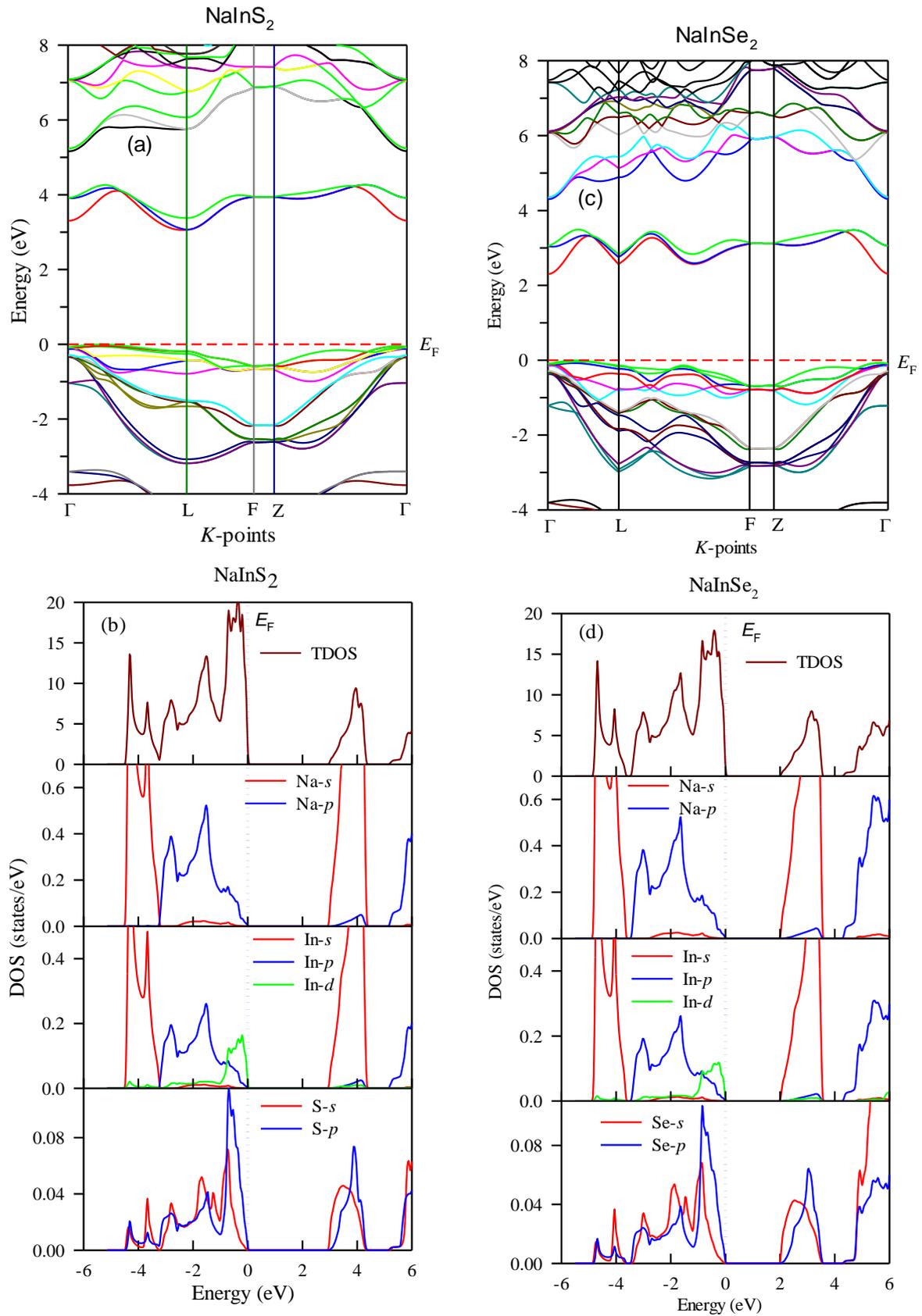

**Fig. 2.** Calculated (a) electronic band structure, (b) density of states for NaInS$_2$, and (c) electronic band structure, (d) density of states for NaInSe$_2$ compounds obtained using TB-mBJ potential.

On the other hand, the bottom of the conduction bands above the $E_F$ is due to the hybridization among $s$ orbital of Na, In and S/Se states. Thus the band containing the CBM has the antibonding character between S/Se-$s$ and In-$p$ orbitals in both thecompounds.This feature is more pronounced in the case of NaInS$_2$ than in NaInSe$_2$ compound [30]. The low energy part of the conduction band, therefore, resides at much higher energy (~ 3 eV) for NaInS$_2$ than (~ 2 eV) for NaInSe$_2$ compound as shown in Figs. 2 (c) and (d). This feature is mainly responsible for the reduced bandgap of the NaInSe$_2$. However, the top part of the valence bands is flatter than the conduction bands for both compounds. The flat nature of these bands comes from the strong hybridization of Na-2$p$, S-3$s$, S/Se-$p$, In-5$p$ and In-4$d$ orbitals which can also be clearly seen from the projected density of states shown in Figs. 2 (c) and (d).For the consideration of electrical conductivity, the In-$d$orbital for the titled compounds should be the dominant contributor to the electronic energy density of states.

**Table 2 -** Calculated Mulliken atomic population for NaInS$_2$ and NaInSe$_2$ compound.

| Compound | Atoms | Mulliken atomic population | | | | | |
| --- | --- | --- | --- | --- | --- | --- | --- |
| | | $s$ | $p$ | $d$ | Total | Charge (e) | EVC (e) |
| NaInS$_2$ | Na | 2.18 | 6.22 | --- | 8.39 | 0.61 | 0.39 |
| | In | 1.01 | 1.32 | 10.00 | 12.33 | 0.67 | 2.33 |
| | S | 1.87 | 4.77 | --- | 6.64 | -0.64 | --- |
| NaInSe$_2$ | Na | 2.22 | 6.38 | --- | 8.60 | 0.40 | 0.60 |
| | In | 1.33 | 1.42 | 10.00 | 12.75 | 0.25 | 2.75 |
| | Se | 1.65 | 4.67 | --- | 6.33 | -0.33 | --- |

The Mulliken atomic population (MAP) analysis and the charge density mapping (CDM) are also carried out with the help of the CASTEP code [15] to gain knowledge on charge transfer mechanism as well as bonding between different atomic species and their nature [31]. The difference between the formal ionic charge and the Mulliken charge on the anion species in the material results in the effective valence charge (EVC). It is seen from the Table 2 that in both compounds, the charge of S/Se is negative; therefore, these elements should accept electron from other species in the material. Here, the MAP analysis revealed that in compound NaInS$_2$, the species In and/or Na donate 0.64 |$e$| charge to S while those species transfer 0.33 |$e$| charge to Se in the NaInSe$_2$ compound[11]. Therefore, the chemical bond could be formed only between Na-S/Se and In-S/Se due to the negative charge on S/Se. It is also fair to note that the bond between In-S/Se could be more covalent than Na-S/Se bond as the higher value of EVC signifies the higher level of covalency.

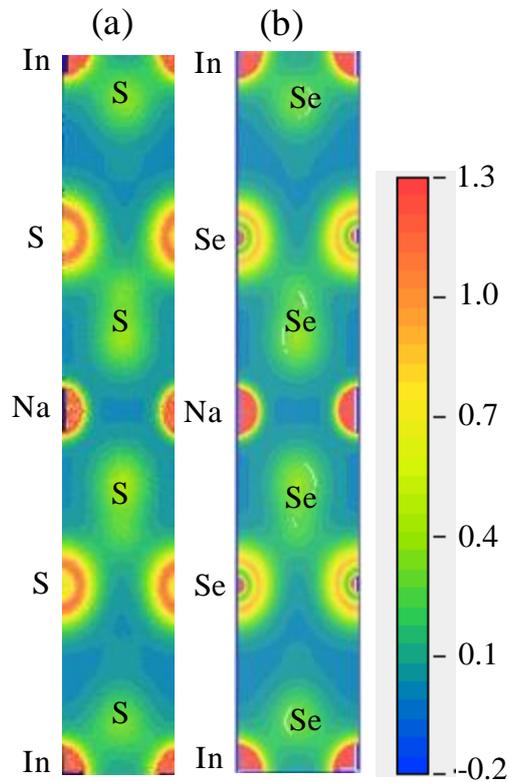

**Fig. 3.** Mapping of electronic charge densities: (a) for NaInS$_2$ and (b) for NaInSe$_2$ compound.

The CDM has also been studied to know the electron densities involved in the chemical bonds between the different atomic species. Fig. 3 (a) and (b) display the CDM for NaInS$_2$ and NaInSe$_2$ compounds. A strong accumulation of charge is found around In and Na atoms whereas strong depletion of charge is seen around S/Se atoms [32]. As observed in Fig. 3, strong covalent bonding can be attributed between In and S atoms whereas comparatively weaker covalent bond is found between Na and S atom for the NaInS$_2$ compound. Similar bonding character could also be observed in the case of NaInSe$_2$ compound. These phenomena are consistent with the MAP analysis.

It is known that thermoelectric properties of materials are directly related to their electronic structure. The favorable band gap of the titled compounds has encouraged us to perform calculations regarding thermoelectric properties, which is described in the next section.

3.3 Thermoelectric transport properties

Since the study of electronic properties confirms the semiconducting nature of the NaInX$_2$ (X=S, Se), it is interesting to study their thermoelectric properties. A potential thermoelectric material should have low thermal conductivity ($\kappa$) and high power factor ($S^2\sigma$). The high power factor simply depends on the high Seebeck coefficient and electrical conductivity. A suitable combination of these two parameters ensuresa high thermoelectric figure of merit (*ZT*) [33], which is a challenging task to achieve. The temperature-dependent thermoelectric transport properties such as Seebeck coefficient (*S*), the ratio between electrical conductivity and the relaxation time ($\sigma/\tau$), power factor ($S^2\sigma/\tau$) and thermal conductivity ($\kappa_{total}$) have been calculated using TB-mBJ potential along two crystallographic directions, *a* and *c* in the temperature range from 300 to 1000 K. The phonon contribution to the thermal conductivity of the compounds of interest was calculated in a recent study [9]. All these transport coefficients are plotted for the NaInX$_2$ (X=S, Se) compounds in Figs. 4 (a-h). The anisotropy in the electronic band structure leads to the anisotropic features in the transport coefficients. The Seebeck coefficient *S* measures the potential difference in different semiconductors or metals induced between two junctions when a temperature gradient is established. The magnitude of this parameter depends critically on the energy derivative of the logarithmic of the electrical conductivity. Therefore, a very close link with the underlying electronic band structure is a natural consequence. It should be noted here that to calculate the different thermoelectric parameters, a fixed relaxation time $\tau=10^{-14}$s has been used in calculations as for most of the semiconducting materials this is in the range of $10^{-14}$s – $10^{-15}$s [34,35]. It is explicit from Figs. 4(a) and (e) that a gradual increasing trend of *S* along *c* direction is observed with temperature upto 800 K and then it reaches a constant value of 295 μV/K and 300 μV/K for NaInS$_2$ and NaInSe$_2$, respectively. This implies that thermally generated electron-hole pairs contribute positively to *S* along the *c* direction.On the other hand, the value of *S* along *a* direction is gradually decreased with temperature. The positive value of *S* attests the domination of *p*-type conduction in both NaInS$_2$ and NaInSe$_2$.

The very low value of $\sigma/\tau$ at low temperatures represents the near insulating behavior of these compounds (very few of the electrons in the valence band are transferred to the conduction band for charge transport). It is seen from Figs. 4 (b) and (f) that $\sigma/\tau$ increases rapidly with increasing temperature and this behavior indicates the semiconducting nature of NaInS$_2$ and NaInSe$_2$ compounds.

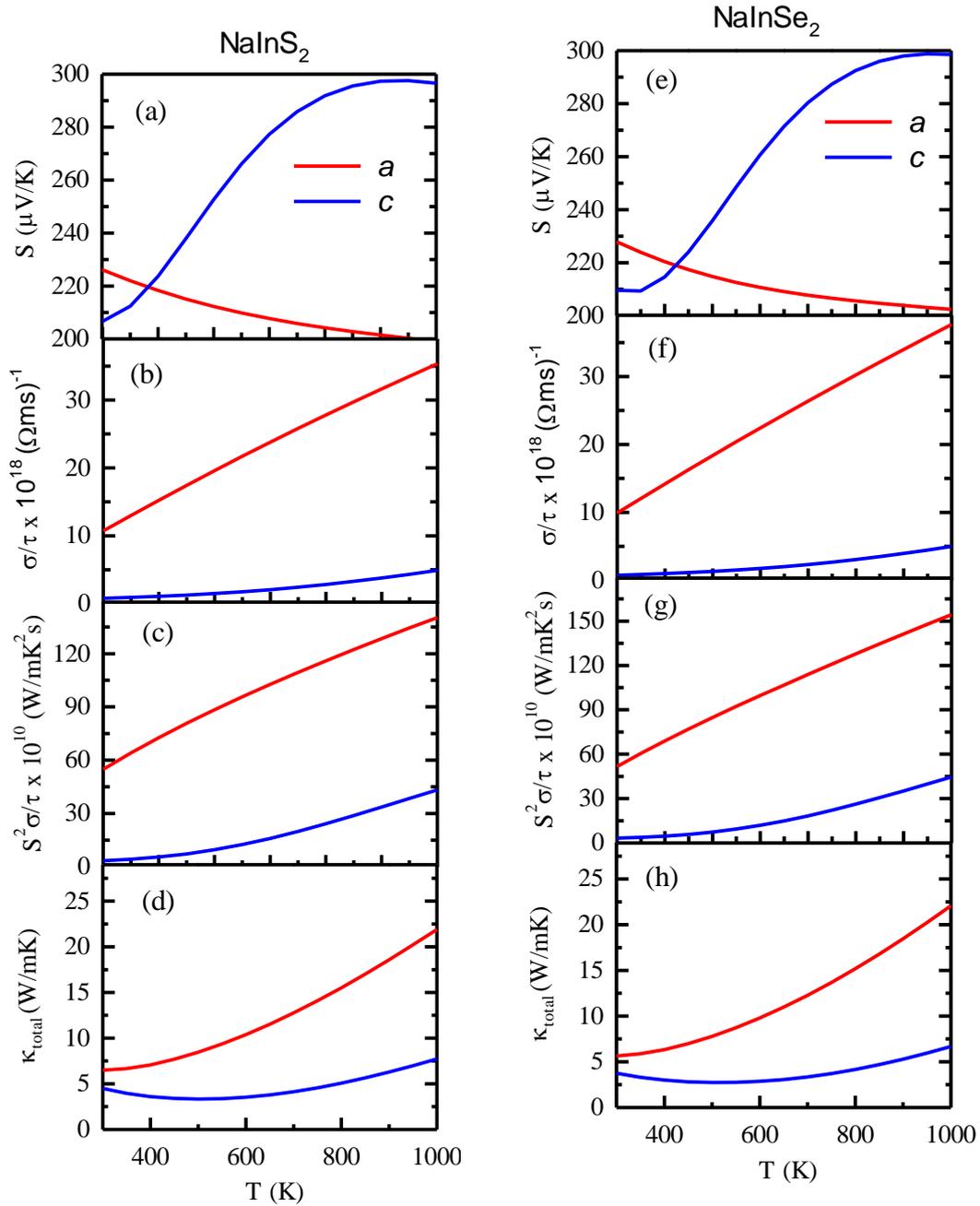

**Fig. 4.** Temperature dependence of (a) Seebeck coefficient ($S$), (b) electrical conductivity over the relaxation time ($\sigma/\tau$), (c) power factor ($S^2\sigma/\tau$) and (d) thermal conductivity ($\kappa_{total}$) along $a$ and $c$ directions for the NaInS$_2$ compound at temperature 900 K. (e-h) the same parameters for the NaInSe$_2$ compound.

The estimated value of $\sigma/\tau$ at room temperature (300 K) is found to be $10.74\times10^{18}$ $(\Omega ms)^{-1}$ and $9.93\times10^{18}$ $(\Omega ms)^{-1}$ along $a$ direction and $0.8065\times10^{18}$ $(\Omega ms)^{-1}$ and $0.7442\times10^{18}$ $(\Omega ms)^{-1}$ along $c$ direction for the NaInS$_2$ and NaInSe$_2$ compounds, respectively. These results clearly suggest that the $\sigma/\tau$ along $a$ direction is much higher than that along $c$ direction. The power factors ($S^2\sigma/\tau$) for both compounds also follow the same trend of $S$ and $\sigma/\tau$ as depicted in Figs. 4 (c) and (g). The maximum estimated values of $S^2\sigma/\tau$ (with $\tau = 10^{-14}$ s) for the NaInS$_2$ at 1000 K are 151.49 μW/cmK$^2$ and 52.77 μW/cmK$^2$ whereas those for the NaInSe$_2$ are

154.34 μW/cmK² and 44.58 μW/cmK² along *a* and *c* directions, respectively. These values for both compounds along *a* direction are much higher (~15 times) than that of 10.1 μW/cmK² along *b*-axis at 850K for the state-of-the-art thermoelectric material, SnSe [14]. The thermal conductivity, $\kappa_{total}$ is a combination of electronic thermal conductivity ($\kappa_e$) and lattice thermal conductivity ($\kappa_l$) where the $\kappa_l$ measures the heat conduction due to the vibration of the lattice ions in a material. The details calculation of $\kappa_l$ for these studied compounds can be found in the literature[9]. The values of $\kappa_{total}$ for the NaInS$_2$ and NaInSe$_2$ at 1000 K are almost the same which are ~ 21.9073 W/mK and ~ 7.7156 W/mK along *a* and *c* crystallographic axes, respectively.

We have also found that at low temperatures (less than 300 K), the $\kappa_l$ is dominating but at a higher temperature (above 300 K) $k_e$ starts to dominate. It is readily seen from the Figs. 4 that all thermoelectric parameters for the NaInS$_2$ and NaInSe$_2$ compounds are significantly anisotropic. All parameter values except *S* along crystallographic *a* direction are significantly higher along *a* than in *c* direction. The efficiency of thermoelectric materials can be gauged using the dimensionless figure of merit, $ZT = \frac{S^2\sigma}{k}T$. The calculated *ZT* values along two crystallographic axes for both the compounds are presented in Figs. 5 (a) & (b). It is found that *ZT* values increase with temperature and there is a strong anisotropy upto 800 K and after that both compounds exhibit less anisotropy. The *ZT* values along *a*-axis for both the compounds at 1000K are almost the same and it is estimated to be ~ 0.7. This *ZT* value is higher than that for recently reported body centered tetragonal (BCT) chalcopyrite material NaInSe$_2$[10]

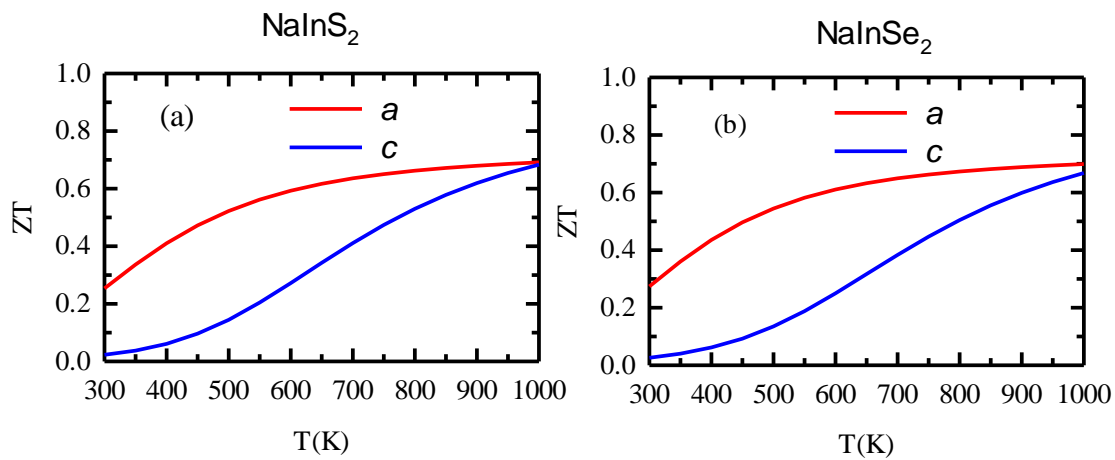

**Fig. 5.** Dimensionless thermoelectric figure of merit (*ZT*) for (a) NaInS$_2$ and (b) NaInSe$_2$ compounds as a function of temperature.

We also studied the variations of $S$, $\sigma/\tau$, $\kappa_e/\tau$ and $S^2\sigma/\tau$ with carrier concentration at 900 K for both materials as depictedin Figs. 6 (a-h). As the carrier concentration increases, $S$ decreases linearly for both compounds as shown in Figs. 6 (a) and (e). It is also clear from Figs. 6 (b) and (f) that a gradually increasing trend of $\sigma/\tau$ for both compounds is observed as a function of carrier concentration.The present thermoelectric materialshave the maximum power factor for carrier concentrations within the range of $10^{21}$ to $10^{22}$ cm$^{-3}$ as shown in Figs. 6(c) and (g). In Figs. 6(d) and (h), we see that the $\kappa_e/\tau$ follows the same trend in accordance with the Wiedemann Franz law; which states that the electronic contribution to the thermal conductivity is proportional to the electrical conductivity. We have summarized the data at temperatures 300 K and 1000 K for different thermoelectric parameters for the NaInS$_2$ and NaInSe$_2$ compounds in Table 3.The obtained $ZT$ values are not suitable for direct practical applications and the high value of $\kappa$ is mainly responsible for this low value of $ZT$. The reduction of thermal conductivity is very important to achieve significantly higher $ZT$ value for practical applications in thermoelectric devices. A strategy (e.g., by nano-structuring [36] and all-scale hierarchical architecture [37]) to reduce lattice thermal conductivity, $k_l$ might significantly increase the efficiency of these prospective thermoelectric materials. The reduction of band gap of both compounds through band engineering will also be interesting and can lead to the enhancement of thermoelectric performance. However, all the other studied thermoelectric parameters, especially high power factor and figure of merit, clearly indicate that both NaInX$_2$ (X=S, Se) compounds have potentials to be used for thermoelectric device applications.

**Table 3** Seebeck coefficient ($S$), electrical conductivity ($\sigma$), power factor ($S^2\sigma$) and thermal conductivity ($\kappa_{total}$) (with $\tau=10^{-14}$ s) along '$a$' and '$c$' directions for NaInX$_2$ (X=S, Se) crystals.

| Compound | Temperature (K) | $S$ ($\mu$V/K) | $\sigma \times 10^4 (\Omega m)^{-1}$ | $(S^2\sigma)_{max}$ ($\mu$W/cmK$^2$) | $\kappa_{total}$ (W/mK) |
|---|---|---|---|---|---|
| NaInS$_2$ | 300 | 226.12[a] <br> 206.73 [c] | 10.74 [a] <br> 0.81 [c] | 54.90 [a] <br> 03.44 [c] | 6.49 [a] <br> 04.50 [c] |
| | 1000 | 00.00 [a] <br> 297.00 [c] | 38.91 [a] <br> 06.18 [c] | 151.49 [a] <br> 52.77 [c] | 21.91 [a] <br> 07.72 [c] |
| NaInSe$_2$ | 300 | 227.92 [a] <br> 209.61 [c] | 9.93 [a] <br> 0.74 [c] | 51.56 [a] <br> 03.27 [c] | 05.63 [a] <br> 03.75 [c] |
| | 1000 | 00.00 [a] <br> 300.00 [c] | 37.66 [a] <br> 05.00 [c] | 154.34 [a] <br> 44.58 [c] | 22.07 [a] <br> 06.67 [c] |

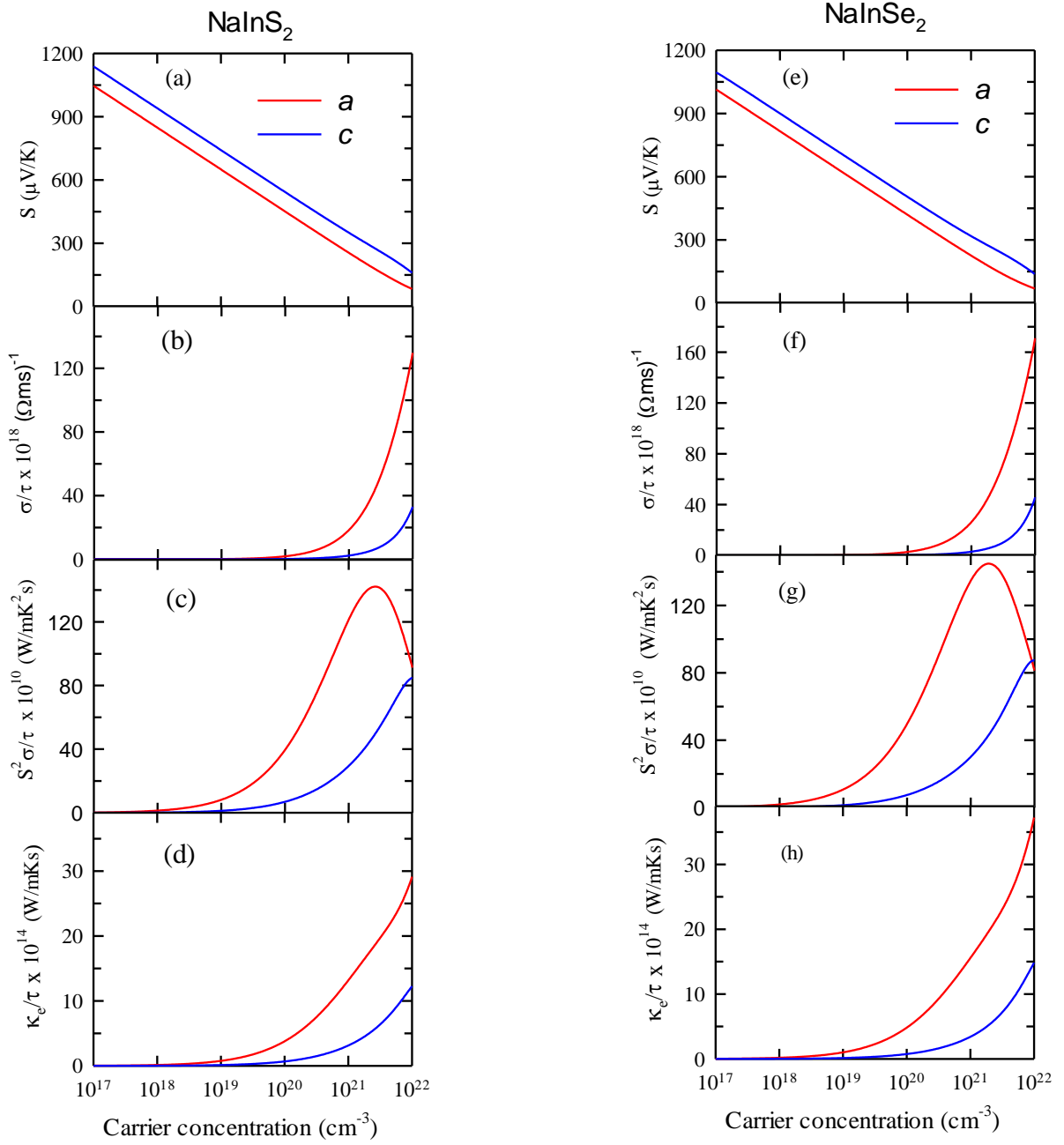

**Fig. 6.** Carrier concentration dependence of (a) Seebeck coefficient ($S$), (b) electrical conductivity over relaxation time ($\sigma/\tau$), (c) power factor ($S^2\sigma/\tau$) and (d) electronic thermal conductivity ($k_e$) along $a$ and $c$ directions for NaInS$_2$ compounds at temperature 900 K. (e-h) same parameters for NaInSe$_2$.

3.4 Optical properties

The optical propertiesof solids are measurable macroscopic physical properties which can be studied by the calculations of the energy dependent dielectric function (directly related to the electronic band structure).The dielectric function (real and imaginary parts) elucidates the various optical properties of solids completely [38]. The study of these properties is also desirable to predict the suitability of solids for possible applications. Therefore, the dielectricfunction $\varepsilon(\omega) = \varepsilon_1(\omega) + i\varepsilon_2(\omega)$ has been calculated and used to study the optical properties of the $NaInS_2$ and $NaInSe_2$ compounds. It is well-known that the imaginary part of the dielectric function, $\varepsilon_2(\omega)$ is a precondition to estimate the rest of the optical constants such as the real part of the dielectric function$\varepsilon_1(\omega)$, refractive index, absorption spectrum, loss-function, reflectivity and optical conductivity of the material. One obtains $\varepsilon_1(\omega)$ from the calculated spectrum of $\varepsilon_2(\omega)$ via the Kramers-Kronig transformation relation [39]. All other optical constants can be extracted from $\varepsilon_1(\omega)$ and $\varepsilon_2(\omega)$. The detail method and related formulae for these optical constantsusing a full-potential augmented plane wave method (FPLAPW) as implemented in WIEN2k can be found in the literature [39,40].

Here, we have studied the optical properties for two polarization directions [100] and [001] and found that the shape of the curves for these polarizations are almost identical but energy positions of various peaks are clearly different and distinguishable. Noted here that the positions of the peak for different optical parameters are gradually shifted to the higher energy region when light is incident onto the materials with [001] polarization as shown in Figs. 7 and 8. Figs. 7 (a) and (e) display the obtained spectra for the imaginary part of the dielectric function which arises simply from direct interband transitions while indirect band transitions are ignored because of their minor contribution due to phonon scattering contribution to the dielectric function[41]. The real part of the dielectric function is also presented in Figs. 7 (b) and (f). The imaginary part of the dielectric function is related to the absorption spectra of the compounds while the real part is related to the polarizability of the materials [41]. From the onset of optical absorption, the obtained bandgap values are ~ 3.3 and ~2.3 eV for $NaInS_2$ and $NaInSe_2$, respectively, which agree very well with those obtained from electronic band structure calculations. Figs. 7 (a) and (e) also demonstrate that the absorption spectrum for $NaInSe_2$ starts at lower energy than that for $NaInS_2$. The difference in the spectra of $\varepsilon_2(\omega)$ for $NaInS_2$ and $NaInSe_2$ is also due to the difference in the bandgap values which decrease when S is replaced by Se.

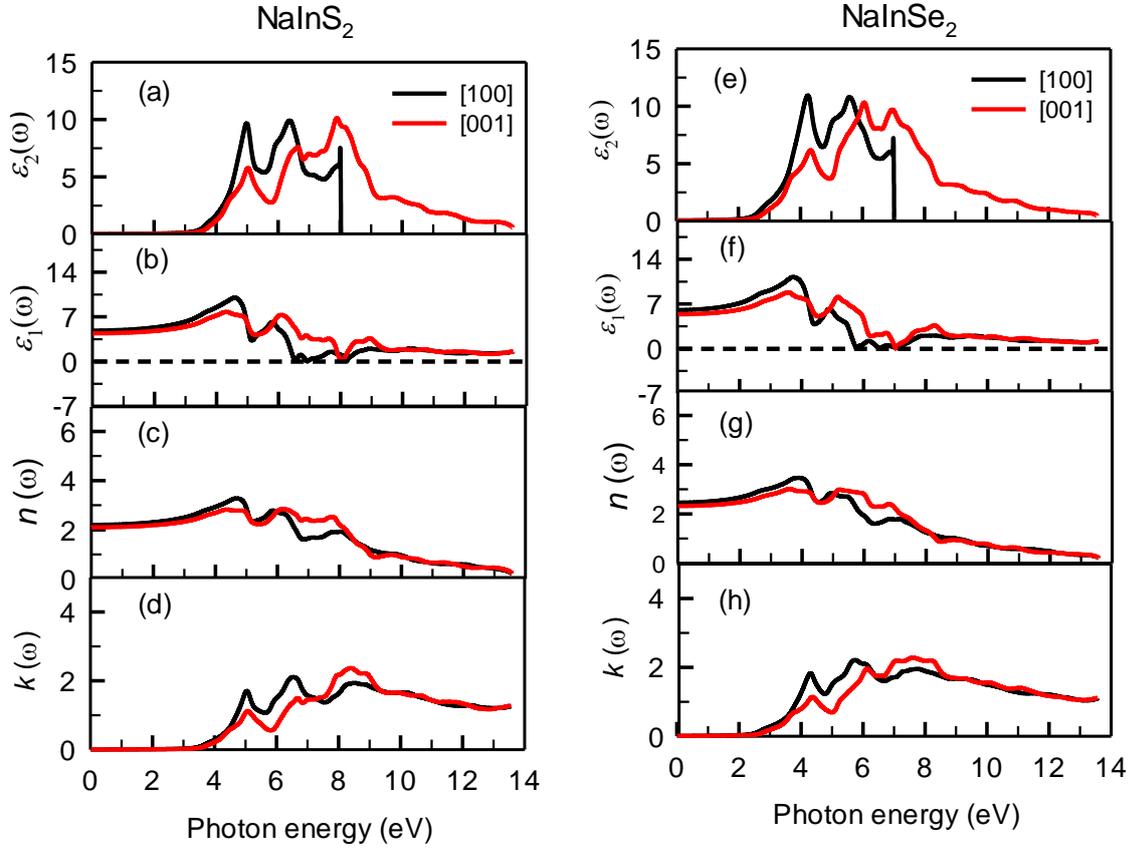

**Fig. 7.** The energy dependence of dielectric function (a) imaginary part, $\varepsilon_2(\omega)$, (b) real part, $\varepsilon_1(\omega)$, (c) refractive index, $n(\omega)$ and (d) extinction coefficient, $k(\omega)$ along [100] and [001] polarization directions for $NaInS_2$ compound. (e-h) indicate same parameters for $NaInSe_2$ compound.

The three consecutive sharp peaks at around 5.0, 6.4, 8.0 and 4.2, 5.6, 7.0 (all in eV) for [100] polarization whereas those at around 5.0, 6.6, 7.8 and 4.4, 6.0, 6.9 (all in eV) for [001] polarization for $NaInS_2$ and $NaInSe_2$, respectively, occur due to the interband transitions from S-$s/p$ and Se-$s/p$ states to the In-$s$ states at the bottom of the conduction bands. The real part of dielectric constant $\varepsilon_1(\omega)$ at zero frequency is known as the static dielectric constant $\varepsilon_1(0)$. The calculated values of the real part of the static dielectric function, $\varepsilon_1(0)$, static refractive index, $n(0)$, absorption band edge (ABE(0)) and reflectivity ($R$) along [100] and [001] polarization directions for $NaInS_2$ and $NaInSe_2$ compounds at zero photon energy are summarized in Table 4. The values of $\varepsilon_1(0)$ are ~ 4.85 and 6.05 for [100] polarization and 4.40 and 5.37 for [001] polarization for $NaInS_2$ and $NaInSe_2$, respectively. These values are in good accord with the Penn model[42]expressed as, $\varepsilon_1(0) = 1 + (\hbar\omega_p/E_g)^2$. It means that solids with high values of $\varepsilon_1(0)$ have lower energy band gaps for a fixed plasma frequency.

The real part of the refractive index, $n(\omega)$ measures the phase velocity whereas the amount of absorption loss is usually measured by the imaginary part (extinction coefficient) when light (photon) penetrates into the material. The obtained $n(\omega)$ of $NaInS_2$ and $NaInSe_2$ are shown in Figs. 7 (c) and (g) where the static refractive index $n(0)$ have the values of ranging 2.1-2.20 and 2.32-2.46 for $NaInS_2$ and $NaInSe_2$, respectively. The value of the refractive index is lower for the semiconductors that have higher bandgap energy. The maximum refractive index (MRI) for $NaInS_2$ and $NaInSe_2$ is obtained at energy of 3.30 eV for [100] polarization and at 3.50 eV for [100] polarization, respectively. The MRI occurs at the corresponding incident energy where the transitions of electrons from the valence band to conduction band occur and the transition is referred to as direct electron transition. Figs. 7(d) and (h) show the variation of the extinction coefficient with photon energy. The extinction coefficient increases with energy and the maximum values are attained at around 8.50 eV and 7.80 eV for [001] polarization for $NaInS_2$ and $NaInSe_2$ compounds, respectively. It is also observed that the variation of $n(\omega)$ and $k(\omega)$ largely follow the variation of $\varepsilon_1(\omega)$ and $\varepsilon_2(\omega)$, respectively, with incident photon energy. It is noteworthy that the values of $\varepsilon_1(0)$ and $n(0)$ of both the compounds under study reveal higher values than those for the tetragonal chalcopyrite $NaInS_2$ and $NaInSe_2$ compounds[10]. It is known that high values of $\varepsilon_1(0)$ and $n(0)$ are useful in optoelectronic device fabrications including LCDs, OLEDs, and quantum dot (QDLED) display.

Figs. 8 (a) and (e) exhibit the optical absorption coefficients that provide information about loss of energy of penetrating light through the solids and prerequisite knowledge to design optoelectronic devices. The spectra started to rise (also known as absorption edge) after a finite value of incident energy approximately equal to the energy band gap values for both compounds owing to their semiconducting nature. It is also noticeable that the spectrum for $NaInSe_2$ starts at lower energy than $NaInS_2$ because of its lower band gap as shown in Table 4. The spectra are observed to increase with some prominent peak with the variation of photon energy. The strongest absorption region is from 3.0 eV to 11.0 eV for $NaInS_2$ and from 2.5 eV to 12 eV for $NaInSe_2$. The highest broad absorption peak in the ultraviolet (UV) region inspires the possible use of these materials to make devices to sterilize surgical equipments [43].

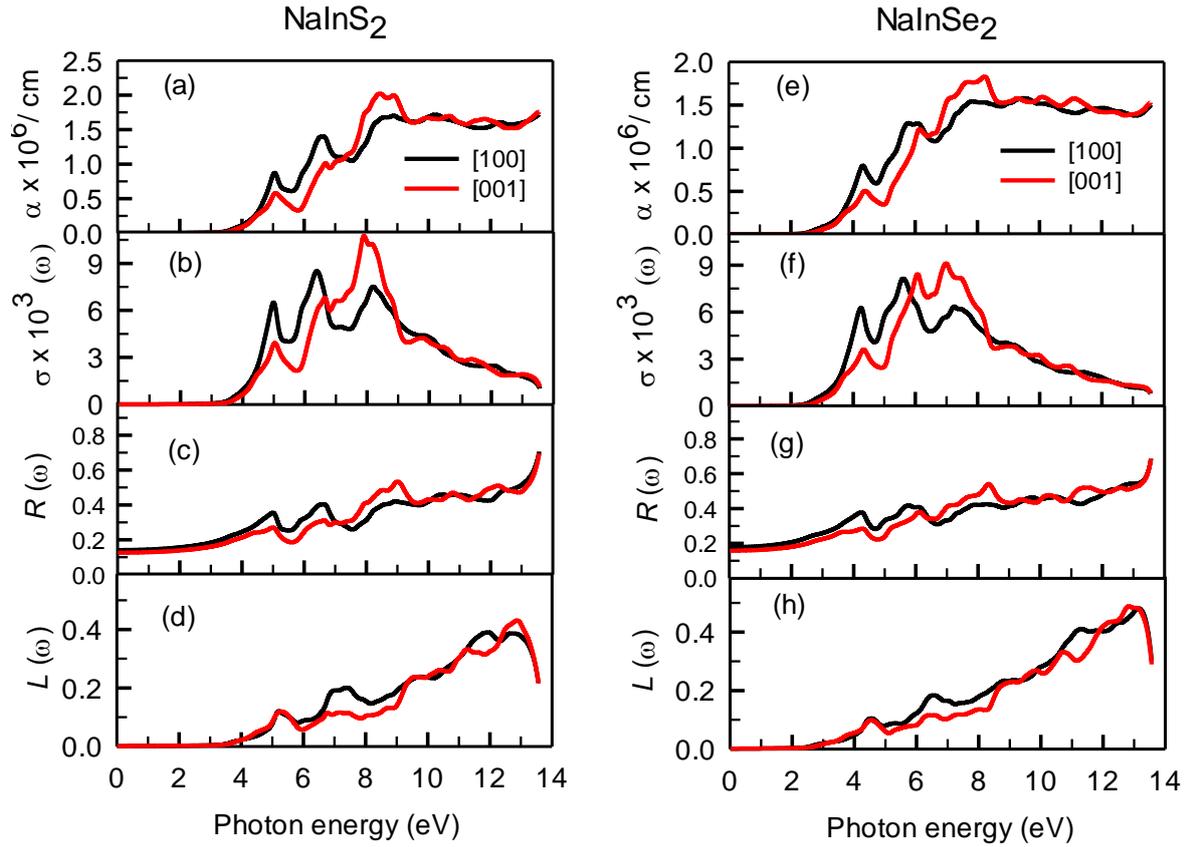

**Fig. 8.** The energy dependence of (a) absorption coefficient, α(ω), (b) photoconductivity, σ(ω), (c) reflectivity, R(ω) and (d) loss function, L(ω) along [100] and [001] polarization direction for $NaInS_2$ compound. (e-h) indicate same parameters for $NaInSe_2$ compound.

Figs. 8(b) and (f) illustrate the photoconductivity spectra characterizing semiconducting nature [44] of the compounds under investigation. The spectral features seen from the absorption coefficient and photoconductivity agree very well with the band structure calculations. The highest values of photoconductivity are found at ~ 8.0 eV and ~ 7.0 eV for [100] polarization for $NaInS_2$ and $NaInSe_2$, respectively.

The photon energy dependent reflectivity spectra of $NaInS_2$ and $NaInSe_2$ are shown in Figs. 8(c) and (g). The value of reflectivity at zero photon energy corresponds to the static part of the reflectivity. The low values of reflectivity in the visible and ultraviolet region make $NaInS_2$ and $NaInSe_2$ potential candidates for use in the area of transparent coatings in the visible and deep UV regions [29]. The reflectivity spectra are noted to increase with photon energy and reached maximum at around 13.0 eV. The loss function (LF) L(ω), is an important optical function, used to describe the energy attenuation of fast moving electrons passing through the material. The plasma frequency ω$_p$ is defined as the frequency of the peak in $L(\omega)$ that also corresponds to the frequency where the reflectivity spectra falls sharply from a high value [45]. The highest sharp peak in the loss function is observed at around 13 eV due to bulk plasmonic excitation for both the compounds, which corresponds to the sudden decline of reflectivity. The energy at this point (maximum of $L(\omega)$) is termed as plasmon energy and the corresponding frequency is called plasma frequency, ω$_p$.

Finally, the tunable electronic band gap, high dielectric constant, good absorption spectra and high photoconductivity clearly indicate that both the ternary layered chalcogenide materials (NaInS$_2$ and NaInSe$_2$) can be used in optoelectronic applications.

**Table 4**- Calculated values of real part of static dielectric function, $\varepsilon_1(0)$, static refractive index, $n(0)$, absorption band edge, ABE(0) and reflectivity ($R$) along [100] and [001] polarization directions for NaInS$_2$ and NaInSe$_2$ compounds at zero photon energy.

| Compound | $\varepsilon_1(0)$ [100] | $\varepsilon_1(0)$ [001] | $n(0)$ [100] | $n(0)$ [001] | ABE (0) [100] (eV) | ABE(0) [100] (eV) | $R(0)\times100$ % [100] | $R(0)\times100$ % [001] |
|---|---|---|---|---|---|---|---|---|
| NaInS$_2$ | 4.85 | 4.41 | 2.20 | 2.10 | 3.01 | 3.20 | 14.10 | 12.59 |
| NaInSe$_2$ | 6.06 | 5.38 | 2.46 | 2.32 | 2.25 | 2.46 | 17.82 | 15.80 |

## 4 Conclusions

To summarize, we have computed the structural and unexplored electronic (density of state, Mulliken atomic population and charge density), optical and thermoelectric transport properties of NaInS$_2$ and NaInSe$_2$ using the first principles method. Our calculated lattice parameters are consistent with the experimentally found results. A distinct band gap between valence and conduction band is observed. The analysis of the Mulliken bond population and charge density mapping shows that the bond between In-S/Se should be more covalent than the Na-S/Se bond. Optical and thermoelectric properties are anisotropic. Carrier concentration dependent thermoelectric properties at 900 K such as Seebeck coefficient ($S$), electrical conductivity ($\sigma$), power factor ($S^2\sigma$) and electronic thermal conductivity ($k_e$) have been calculated and discussed. The highest values of $S$ (in μV/K) of 295 and 300 at temperature 1000 K along *a* direction are realized for NaInS$_2$ and NaInSe$_2$ compounds, respectively. The positive value of $S$ attests the dominance of $p$-type conduction channel. The electrical conductivity ($\sigma$) at 300 K is estimated to be 10.74×10$^4$ ($\Omega$m)$^{-1}$ and 9.93 ×10$^4$ ($\Omega$m)$^{-1}$ along $a$ direction and 0.8065×10$^4$ ($\Omega$m)$^{-1}$ and 0.7442×10$^4$ ($\Omega$m)$^{-1}$ along $c$ direction for NaInS$_2$ and NaInSe$_2$ compounds, respectively. This shows that the $\sigma$ along $a$ direction is much higher than that along $c$ direction. The power factor ($S^2\sigma$) exceeds 151.34 μW/cmK$^2$ for both compounds along $a$ direction which is much higher than that of the state-of-the-art thermoelectric material, SnSe of 10.1 μW/cmK$^2$. The obtained $ZT$ value of ~ 0.7 is higher than that of the tetragonal chalcopyrite of NaInSe$_2$. The $ZT$ value can be further increased via band gap engineering. Another approach could be insertion of suitable atomic layers in NaInS$_2$ and NaInSe$_2$ to reduce the thermal conductivity without affecting the electrical conductivity. Such intercalation is relatively easy for materials which have layered structure to begin with. The absorption coefficient and photoconductivity spectra confirm the semiconducting nature of these materials, which are consistent with calculated electronic band structure and previous reports. The sharp absorption peak in the ultraviolet region indicates the possible use of these materials for fabricating devices to sterilize surgical equipments. Based on impressive optical and thermoelectric properties, it can be concluded that the ternary layered chalcogenides NaInS$_2$ and NaInSe$_2$ could be suitable candidates for thermoelectric and optoelectronic device applications. Experimental verifications of the results presented herein are expected and we hope that this study will serve as a reference for future experimental and theoretical studies on these interesting systems which hold significant prospect in energy harvesting and optical device applications.

Acknowledgements

Authors are grateful to the Department of Physics, Chittagong University of Engineering & Technology (CUET), Chattogram-4349, Bangladesh, for providing the computing facilities for this work.

Data availability

The datasets generated during the current study are available from the corresponding author on a reasonable request.

Conflict of Interest

The authors declare that they have no known competing financial interests or personal relationships that could have appeared to influence the work reported in this paper.